# Distributed Simulation of Large Multi-body Systems

Manas Kale[1,2] and Paul G. Kry[1,2]

[1]McGill University
[2] Huawei Technologies Canada

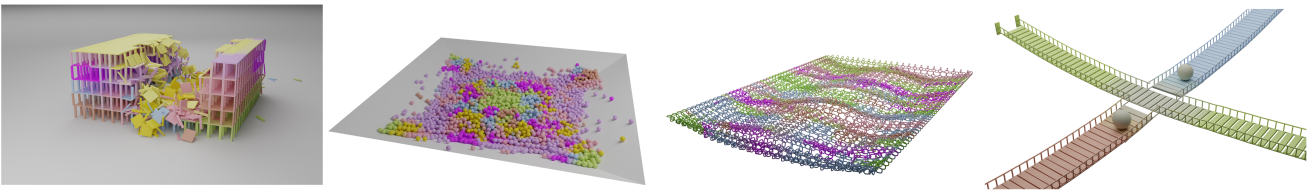

**Figure 1:** *Example simulations where colors show the division of computation across different workers. Left to right: Building, Bowl, Chain, and Four-way Bridge simulated using 8, 7, 6, and 3 workers respectively. Multiple workers simulate the same rigid body within overlap regions, seen above with bodies drawn with blended colors that signify the weights used for computing their blended states. The division of work evolves over time to maintain a balanced load distribution.*

## Abstract

*We present a technique designed for parallelizing large rigid body simulations, capable of exploiting multiple CPU cores within a computer and across a network. Our approach can be applied to simulate both unilateral and bilateral constraints, requiring straightforward modifications to the underlying physics engine. Starting from an approximate partitioning, we identify interface bodies and add them to overlapping sets such that they are simulated by multiple workers. At each timestep, we blend the states of overlap bodies using weights based on graph geodesic distances within the constraint graph. The use of overlap simulation also allows us to perform load balancing using efficient local evaluations of the constraint graph. We demonstrate our technique's scalability and load balancing capabilities using several large-scale scenes.*

## 1. Introduction

Over the past two decades, a convergence of mathematical models and computational power has led to many successful research results for physics-based animation rigid bodies with constraints and frictional contact [BET14]. These methods have been widely used in the gaming, film, robotics, and simulation industries.

We can characterize rigid body simulations as having three desirable properties: performance, robustness, and accuracy. Better performance allows larger or faster simulations, better accuracy allows greater prediction of object motion and better robustness allows the simulation to behave gracefully in complex situations. Although there has been much research on improving the accuracy and robustness [TBV12, PAK*19], recent trends in the industry have focused on delivering faster and larger simulations (i.e., performance), motivated by large-scale movie and video game pro-ductions. In this work, our primary focus lies in addressing the performance aspect of rigid body simulations.

A significant opportunity for performance improvement is to use increasingly available on-demand cloud computing technologies. By horizontally scaling simulation workloads across heterogeneous CPUs, we can harness the scalability of the cloud to run large-scale simulations. An important challenge here is *maintaining* balanced workload distribution in a simulation that is fundamentally chaotic and influenced by unpredictable user input. For example, it difficult to decide partition assignments for a collapsing pile of bodies as the behavior is unpredictable.

In particular, the issue involves handling collisions between interface bodies, i.e., collisions between bodies that are in different partitions. If a body in one partition collides with a stack of bodies in another, having a rule that transfers one body to the other partition will eventually lead to an imbalance. Likewise, dividing the 3D simulation world into a per-partition grid may also lead to load imbalance if all bodies are concentrated in a single grid cell, as is the case in interactive games. Optimal load-balancing in such situations requires knowledge of how the constraint graph is going to evolve, but computing this requires running the simulation itself.

We address this challenge by using an overlap algorithm that simulates interface bodies in multiple overlapping sets and *blends* their results. Using overlapping sets creates opportunities for the algorithm to drive the system towards a more balanced load distribution by using simple greedy heuristics relying solely on local evaluations of the constraint graph topology. The same algorithm is also used to simulate large articulated bodies in different overlapping sets. Consequently, our technique supports parallel simulation of both freely moving and articulated bodies, suitable for distributed computing while maintaining a balanced workload. Our method is similar in spirit to recent works that manipulate effective mass [TBV12, PAK*19] but is simpler in the sense that it works at the rigid body level, requiring no modifications to the underlying solver. We manipulate only rigid transformations and velocities to achieve the blending of interface bodies, making our method easy to adapt to existing physics engines using simple API modifications. In summary, our method

- exploits overlapping sets of bodies for parallel computation;
- couples bodies using constraint graph geodesic distances;
- provides load balancing for contact-rich scenes;
- and is agnostic to the underlying solver.

## 2. Related Work

This work is inspired by a diverse body of previous work that spans physically-based simulation methods, graph partitioning, and distributed computation. Here we provide a brief overview of how our method compares to related methods.

*Constrained Multibody Dynamics.* The treatment of constraint dynamics as mixed linear complementarity problems (MLCPs) is widespread in the graphics community [ST96, TNA08, ELA19]. Complementarity problems arising in rigid body systems can be solved either iteratively or directly, and the choice between these methods hinges on the trade-off between speed and accuracy. Projected Gauss-Seidel is a widely used solver [Erl05, Cat05] with limited parallelism capabilities because updates to bodies with multiple constraints have to be serialized. Jacobi-based solvers are capable of parallelization but at the cost of slower convergence speed. Tonge et al. [TBV12] describe a parallel projected Jacobi-based method to prevent jitters in large piles of bodies, using the novel idea of mass splitting along with a graph coloring algorithm. The mass term for each body is divided by its contact count, but the full mass is used for applying impulses. This is similar to our method in the sense that our method also divides the effective mass among workers by using different weights for velocities and rigid transformations.

For multi-body systems, a common way to parallelize is to divide the system into subsystems connected at interfaces and solve them in parallel. Featherstone [Fea99] proposes such a divide-and-conquer algorithm for systems with bilateral constraints. The algorithm arranges constraints as a tree and solves them in a bottom-up manner, children then root nodes. Other methods modify this algorithm to work both on the CPU and GPU [CAB09, LAKP14]. We note that this divide-and-conquer formulation is primarily targeted at articulated systems with joint constraints, whereas contacts can introduce the additional challenges of being intermittent and the formation of loops in the constraint graph.

Peiret et al. [PAK*19] propose a method that handles both unilateral and bilateral constraints, targeting stiff systems using direct solvers. Here, after subsystems are split and solved separately, they are coupled using reduced-order models using a method based on the Schur complement domain decomposition. The use of semantic partitioning described in their work is used as a starting point in our algorithm. However, the major distinction is that while their work uses non-overlapping domains and tight coupling constraints, our approach uses overlapping domains to achieve coupling while also permitting an algorithm for dividing the work in a balanced manner through local evaluations of the constraint graph.

*Partitioning.* For large graphs, such as large-scale rigid body simulations, general graph partitioning algorithms such as METIS [KK98] are known to provide excellent results. However, in simulations where coupling can change due to contact it can be costly to use this method for partitioning at every time step. Partitioning algorithms for dynamic graphs must either minimize the need to evaluate the whole constraint graph or perform partitioning in parallel. The work by Liu and Andrews [LA22] is an example of the first category. Here, various graph partitioning algorithms are explored to produce non-overlapping subsystems for parallel simulation. The proposed algorithms focus on contact-rich simulations, using heuristics to avoid creating poorly conditioned partitions and reduce edges connecting adjacent partitions. Notably, the need for avoiding a global evaluation of the constraint graph is observed, which is something we also do in our work.

Partitioning schemes in soft-body simulations is another active area of research. In contrast to breaking work into connected blocks, graph coloring-based methods are common for decoupling the degrees of freedom in a manner that permits a parallel Gauss-Seidel solve. For example, Fratarcangeli and Pellacini [FP15] partition position based dynamics soft body simulations [MHHR07] using the dual of the constraint graph, where graph nodes are constraints and edges are particles. Further work proposes a randomized graph coloring at the beginning of each frame permitting the use of a parallel randomized Gauss-Seidel solver [FTP16]. Since graph coloring involves a significant overhead, they use a parallelizable graph coloring approach. More recent work by Ton-That et al. [TTKA23] identifies highly dependent nodes to create supernodes, reducing partitions and improving performance. The use of graph coloring for soft bodies and piles of rigid bodies [TBV12] emphasizes the need for minimizing the number of partitions, thereby reducing the number of interface nodes. In our work, we do not use graph coloring, and furthermore, instead of a dynamic or optimized number of partitions, we use a user-specified number of workers to study how our algorithm behaves.

*Distributed Systems.* Notable examples of distributing physics simulations that use cluster computing include recent work on fluid simulations. Fluids are an excellent candidate for distributed computing because they can involve extremely large systems, and therefore scaling provides a substantial benefit [MSQ*18, QMSL20, SHQL18]. Because of the high amount of coupling in such simulations, methods to divide the problem usually involve spatial partitioning as opposed to graph-based methods. Shah et al. [SHQL18] describe a method to load-balance fluid simulations by predicting future load using a concurrent low-resolution simulation. Mashayekhi et al. [MSQ*18] describe a fault-tolerant system for fluid simulations called Nimbus that uses a spatial grid-based division strategy. Notably, the use of state shared by multiple simulations, called ghost regions, is similar to our use of overlapping bodies: ghost regions (i.e., overlap) provide a mechanism to stitch together independent simulations and their size controls the tradeoff between parallelism and performance. For load-balancing, Nimbus periodically evaluates the load on worker nodes as opposed to our method which only performs an evaluation when an inter-partition contact takes place.

Finally, Brown et al. [BUM19] propose a method to distribute contact-rich rigid body simulations in the cloud using a spatial grid-based division strategy. Their main contribution is the ability for bodies to interact across spatial boundaries, achieved through the use of speculative migration techniques. Each body projects an aura around it that allows seamless transitions across boundaries, should such an event actually happen. Our method in contrast uses graph geodesic distance based partitioning, works for both contact and joint constraints, and uses migrations as opportunities to balance load.

## 3. Background

The dominant step in a rigid body simulation loop is the constraint solve, which is what we parallelize in this work. At every timestep, this constraint solve is used to calculate forces that enforce constraints such as contacts and joints. Below we briefly describe the problem, and we refer the reader other sources [AEF22, BET14] for a comprehensive overview of the mathematical models, algorithms, and numerical methods used in rigid body simulations.

The simulation of $n$ bodies and $m$ constraints using a discrete timestep $h$, involves solving first for velocities and constraint forces, and then updating positions with the new velocities. The relation between velocities and forces is given by

$$\mathbf{M}(\mathbf{v}^+ - \mathbf{v}) = h\mathbf{f} + \mathbf{J}^T\lambda^+, \tag{1}$$

where the superscript $^+$ indicates values calculated at the end of the timestep, $\mathbf{v} \in \mathbb{R}^{6n}$ is the generalized velocity vector, $\mathbf{M} \in \mathbb{R}^{6n \times 6n}$ is the symmetric positive definite mass matrix, $\mathbf{J} \in \mathbb{R}^{m \times 6n}$ is the constraint Jacobian matrix, $\mathbf{f} \in \mathbb{R}^{6n}$ the generalized external force vector, and $\lambda^+ \in \mathbb{R}^m$ the unknown constraint impulse vector. To permit a solution to both the velocity and constraint impulse vector, a second equation is used to constrain the velocities. Including regularization to prevent singularities from redundant constraints we have

$$\mathbf{J}\mathbf{v}^+ + \mathbf{C}\lambda^+ + h^{-1}\phi = \mathbf{J}\mathbf{v}, \tag{2}$$

where $\mathbf{C} \in \mathbb{R}^{m \times m}$ is the compliance matrix (often called *constraint force mixing*) and $\phi \in \mathbb{R}^m$ the position-level constraint violation vector which is included as a stabilization term to reduce drift.

Combining Equations 1 and 2 we obtain the linear system

$$\begin{bmatrix} \mathbf{M} & -\mathbf{J}^T \\ \mathbf{J} & \mathbf{C} \end{bmatrix} \begin{bmatrix} \mathbf{v}^+ \\ \lambda^+ \end{bmatrix} = \begin{bmatrix} \mathbf{p} \\ -\mathbf{d} \end{bmatrix}, \tag{3}$$

where momentum $\mathbf{p} = \mathbf{M}\mathbf{v} + h\mathbf{f}$ and $-\mathbf{d} = h^{-1}\phi - \mathbf{J}\mathbf{v}$.

A further optimization is to eliminate $\mathbf{v}^+$ from the system by taking the Schur complement of the block $\mathbf{M}$,

$$\underbrace{(\mathbf{J}\mathbf{M}^{-1}\mathbf{J}^T + \mathbf{C})}_{\mathbf{A}}\lambda^+ = \underbrace{-\mathbf{d} - \mathbf{J}\mathbf{M}^{-1}\mathbf{p}}_{\mathbf{b}}, \tag{4}$$

where the block diagonal form of $\mathbf{M}$ makes it trivial to calculate its inverse. Here we solve for the vector $\lambda^+$ which contains *Lagrange multipliers* that enforce constraints in the system.

In the absence of unilateral constraints, such as frictional contact, Equation 4 is a simple linear solve. In the presence of contacts, however, the constraint velocity $\mathbf{w} = \mathbf{A}\lambda^+ - \mathbf{b}$ is not always zero because frictional contacts must be allowed to slide or separate. The contact problem is often formulated as a linear complementarity problem (LCP), and can also be viewed as a MLCP in the presence of a mix of bilateral and ulilateral constraints, or a mix of upper and lower bounds as is common for approximating friction. In the latter case, the MLCP formulation enforces limits and complementarity conditions on the system such that

$$\mathbf{w} = \mathbf{w}_+ - \mathbf{w}_-, \text{ with } \lambda_{lo} \leq \lambda^+ \leq \lambda_{hi} \quad \text{and} \tag{5}$$

$$0 \leq \mathbf{w}_+ \perp \lambda^+ - \lambda_{lo} \geq 0 \tag{6}$$

$$0 \leq \mathbf{w}_- \perp \lambda_{hi} - \lambda^+ \geq 0, \tag{7}$$

where the constraint velocity $\mathbf{w}$ is split into non-negative complementary components $\mathbf{w}_+$ and $\mathbf{w}_-$. The symbol $\perp$ represents element-wise complementarity and the vectors $\lambda_{hi}$ and $\lambda_{lo}$ are upper and lower limits on the constraint impulses respectively. These limits are different for various types of constraints. For example, the limits of a contact using the Coulomb friction model are computed using the material's coefficient of friction whereas for joints these are usually set to $\pm\infty$.

Equations 5-7 represent the MLCP subsystem that is solved in parallel in our work. A variety of methods can be used to solve it, ranging from iterative methods [Erl07] that continuously refine an approximate solution to direct methods [JP94] that compute a solution within a finite number of steps. Likewise other friction formulations can be used to solve exact friction cones as nonlinear complementarity problems, which may generally be more costly to solve.

In this work, we are agnostic to the underlying formulation and solver because we only manipulate positions, velocities, and orientations, leaving the actual constraint solve to the individual workers. This allows our method to be easily implemented in any physics engine using simple API primitives that provide access to these quantities. We now describe our method in a top-down fashion, starting with the system architecture followed by the underlying algorithms.

## 4. System Architecture

We begin by providing some necessary definitions around grouping participating entities (rigid bodies and constraints). The set $\mathcal{A}$ con-tains all non-static bodies in the simulation. A static body can be seen as having infinite mass (such as the ground) and thus cannot transfer any forces. For instance, a ball falling on a static ground plane has no effect on nearby bodies already on the plane. In Figure 2, we can see an example where $\mathcal{A} = \{1, 2, \cdots 10\}$.

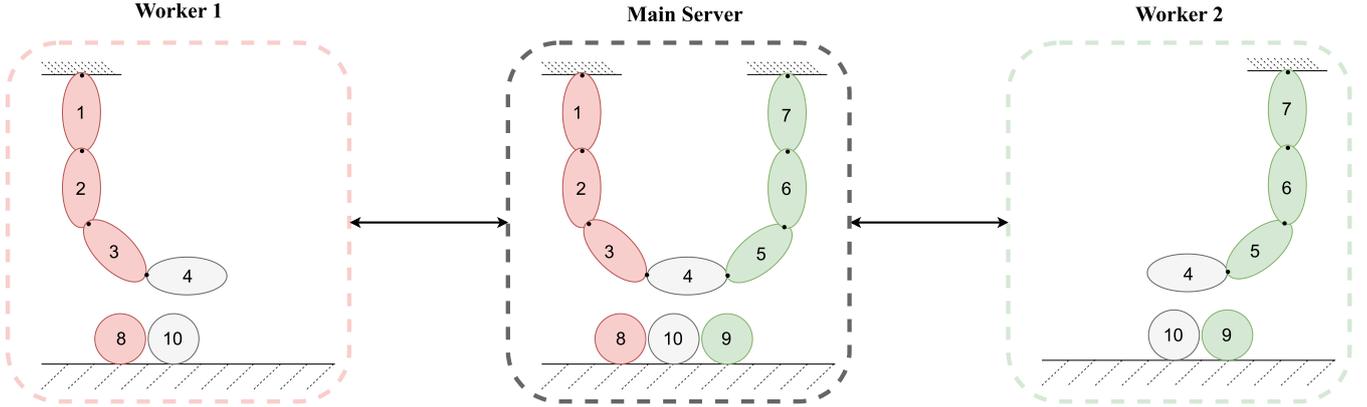

**Figure 2:** *System diagram. Workers contain their subsets of bodies ($\mathcal{P}_1, \mathcal{P}_2, ..., \mathcal{P}_N$) and the main server has a view of all bodies in the simulation ($\mathcal{A}$). Colored bodies exist in a single worker whereas grey bodies are shared between workers.*

We further define the term *active* body to mean a rigid body that is being simulated, as opposed to an *inactive* body. In the rigid body simulation loop, active bodies are evaluated for collision detection, and any constraints involving these bodies are solved. Other than occupying memory, inactive bodies and associated constraints do not participate in the simulation. Note that this is different from *sleeping* [Erl05], a popular technique used in physics engines where bodies are removed from the constraint solve but still participate in collision detection.

We define the generalized state of a rigid body $b$ as

$$\mathbf{s}_b = (\mathbf{x}, \mathbf{q}, \dot{\mathbf{x}}, \boldsymbol{\omega}) \in \mathbb{R}^{13}, \tag{8}$$

where $\mathbf{x}, \mathbf{q}, \dot{\mathbf{x}}, \boldsymbol{\omega}$ are the position, rotation quaternion, linear velocity, and angular velocity respectively in the world frame. We use $\mathbf{s}_b$ to calculate the body's blended state (described later), which is subsequently used as a starting state in the next timestep.

*Workers.* We split our system into $N$ workers (decided before the start of simulation), which are essentially operating system processes running a physics engine instance. At any timestep, each worker has a set of active bodies that it is responsible for simulating. More formally, the set $\mathcal{P}_i$ denotes the subset of active bodies in worker $i$ and is defined as

$$\mathcal{P}_i = \{ p \mid p \in \mathcal{A} \}, \mathcal{P}_i \subseteq \mathcal{A}, \ i = \{1, \cdots N\}, \tag{9}$$

where the subscript $i$ denotes worker index. In Figure 2 $N = 2$, $\mathcal{P}_1 = \{1, 2, 3, 4, 8, 10\}$ and $\mathcal{P}_2 = \{4, 5, 6, 7, 9, 10\}$. Note that some workers may share bodies - we refer to these as *overlap* bodies and this is a central idea of this work. The overlap set $\mathcal{O}$ contains all shared bodies and is given by

$$\mathcal{O} = \{ o \mid o \in \mathcal{P}_i \cap \mathcal{P}_j \, \forall \, i, j = \{1, \cdots N\}, \ i \neq j \}, \tag{10}$$

where $i$ and $j$ are worker indices. In Figure 2, $\mathcal{O} = \{4, 10\}$ and these bodies are shared between workers 1 and 2. At the start, we load all bodies involved in the simulation, both $\mathcal{A}$ and the static bodies, into each worker's memory and the algorithms described later handle the activation and deactivation of these bodies.

*Main Server.* The main server has a global view of the whole simulation as it contains all bodies in $\mathcal{A}$. It is used to assign overlapping sets, run the overlap algorithm, coordinate worker timestepping, and perform state blending. Communication between workers and the main server takes place in the form of commands or data: workers may be instructed to activate or deactivate certain bodies, take a timestep, or be asked to provide a body's generalized state. An important task in our problem is to identify and handle inter-worker contacts. Due to its global view, the main server performs this task, detailed in Section 5.4.

This setup allows our system to be agnostic to the physical location of workers - they can exist on the same or different processing cores as the main server, or even across a network in cloud computing infrastructure. Similar to Baraff and Witkin's [BW97] "black box" approach, each worker exists in isolation and from its point of view, it is only solving for the bodies assigned to it and has no idea about any other worker. The only interaction a worker has with the outside world is when the state of its overlap bodies is reset at the beginning of a main timestep. Thus all workers can perform a timestep (i.e., collision detection, constraint resolution, time integration) for their overlap set in parallel when instructed by the main server. A summary of all frequently used notations in this work is given in Table 1.

## 5. Overlap Algorithm

Let us consider the simple case of an articulated chain hanging from the ceiling at both ends (Figure 3) as an example to describe the method. Each rigid body is connected by a ball-and-socket joint, allowing only rotational degrees of freedom. Our goal is to split this simulation across different workers so that each worker has an approximately equal load. The simplest method to do this would be to assign half of the bodies and joints to two workers as shown in Figure 3(a). However, such partitioning obviously cannot work as the red worker will have no idea about the green portion and vice-versa. From the red worker's point of view, only the bodies $\{1,2,3\}$ and the joints between them exist. If we let this simulation run, each half will behave like a hanging chain that is free at the bottom end.

**Table 1:** *Summary of frequently used notations.*

| Symbol | Description |
|---|---|
| $\mathcal{A}$ | Set of all non-static bodies |
| $\mathcal{P}_i$ | Set of active bodies in a worker $i$ |
| $\mathcal{O}$ | Overlap set |
| $\mathbf{s}_b$ | State vector for a body $b$ |
| $\mathcal{G}$ | Constraint graph |
| $\mathcal{W}_x$ | Set of workers in which body $x$ is active |
| $|\mathcal{W}_x|$ | No. of workers in which body $x$ is active |
| $\langle \mathcal{W}_x \rangle$ | Total number of active bodies in $\mathcal{W}_x$ |
| $activate(\mathcal{W}_x, b)$ | Activate body $b$ in workers $\mathcal{W}_x$ |
| $deactivate(\mathcal{W}_x, b)$ | Deactivate body $b$ in workers $\mathcal{W}_x$ |
| $\gamma$ | Overlap growth depth parameter |

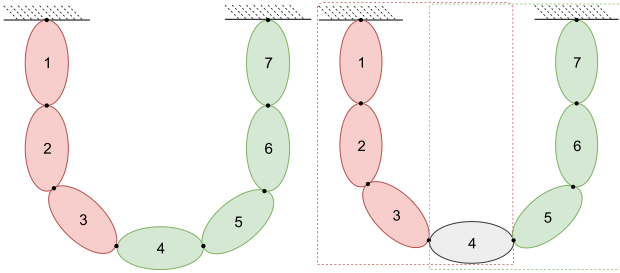

**Figure 3:** *Partitioning for a simple chain hanging from the ceiling. Left: approximate partitioning with interface bodies {3,4}. Right: body 4 is added to the red worker, allowing interface forces to be transmitted across workers.*

Instead of introducing ghost forces, we employ the overlap algorithm to reconcile forces at the interface bodies. Our key insight is to simulate the overlapping body in *both* workers and then *blend* their states to obtain an average state. First, both workers are instructed to take a timestep in parallel. This results in two states being calculated for body 4: $\mathbf{s}^t_{red}$ and $\mathbf{s}^t_{green}$. These two states are then blended together to obtain $\mathbf{s}^t_{blend}$. Finally, for the next timestep $t+1$, the state of body 4 in both workers is reset to the previous blended state $\mathbf{s}^t_{blend}$, and a new simulation iteration begins. Resetting the state before each timestep ensures the two states never diverge by more than a single timestep. Thus their divergence is maximally bounded by the timestep size. This procedure further has the benefit of capturing any partition-internal dynamics acting on the body in the form of a simple blended state. We note that although this algorithm is not physically correct, it does provide visually plausible results.

Further methods described in this paper are built around this fundamental idea. Specifically, we describe a method to assign bodies to the overlap set for any articulated body (Section 5.1), the blending algorithm (Section 5.2), a procedure to calculate blending

---

**Algorithm 1** Overlap algorithm.

**Input:** $\mathcal{G}$ constraint graph of scene with some initial partitioning
**Input:** $\gamma$ overlap growth depth
1: **function** SIMULATE($\gamma, \mathcal{G}$)
2:     ASSIGNTOOVERLAPSET($\gamma, \mathcal{G}$)      ▷ §5.1, Alg. 2
3:     **for all** $o \in \mathcal{O}$ **do**
4:        $\mathbf{w}_o \leftarrow$ COMPUTEWEIGHTS($o, \mathcal{G}$)     ▷ §5.3, Alg. 5
5:     **end for**
6:     **while** not quit **do**
7:        **for all** $o \in \mathcal{O}$ **do**
8:           resetStateTo($\mathcal{W}_o, o, \mathbf{s}^{t-1}_o$)
9:        **end for**
10:      $c^t \leftarrow$ getContacts()
11:      $\mathcal{G}^t \leftarrow$ updateConstraintGraph($c^t$)
12:      LOADBALANCE($c^t, \gamma, \mathcal{G}^t$)     ▷ §5.4, Alg. 6
13:      **for all** $o \in \mathcal{O}$ **do**
14:        $\mathbf{w}^t_o \leftarrow$ COMPUTEWEIGHTS($o, \mathcal{G}^t$)     ▷ §5.3, Alg. 5
15:      **end for**
16:      stepWorkers()        ▷ In parallel
17:      barrier()
18:      **for all** $o \in \mathcal{O}$ **do**
19:        $\mathbf{s}^t_o \leftarrow$ BLEND($o, \mathbf{w}^t_o$)     ▷ §5.2, Alg. 4
20:      **end for**
21:     **end while**
22: **end function**

weights (Section 5.3) and a dynamic load-balancing procedure to handle collisions (Section 5.4) for freely moving bodies. Algorithm 1 describes how these procedures are used in the simulation loop executed on the main server. First, as a preprocessing step, articulated bodies are partitioned and their weights are computed in lines 2-5. Then in the simulation loop, all overlap body states are reset to their previous blended states in lines 7-9. Next, load balancing and weight computation are performed for new contacts in lines 10-15. Workers then take a timestep and solve their subsystems in parallel in line 16. A barrier in line 17 ensures all workers have calculated new states for before proceeding to the next step. Finally, new blended states are calculated for bodies in the overlap set in lines 18-20.

### 5.1. Articulated System Partitioning

*Constraint Graph.* Extending the contact graph formulation described by Erleben [ED04], we define the constraint graph $\mathcal{G}$ as an undirected graph where vertices are rigid bodies, and two vertices are connected by an edge if there exists a constraint between them (joint or contact). We use this constraint graph to perform various operations before and during simulation. Note that static bodies (i.e., bodies with infinite mass) such as the ground are not used here since they are incapable of transmitting any forces.

Using a constraint graph, the problem of partitioning bodies is converted to a graph partitioning problem with the added novel complexity of forming overlap sets. Although well-studied in the context of social networks and high-performance computing, graph partitioning is an NP-Hard problem [GJ90]. We also note a key differentiating characteristic of our constraint graph: certain ver-

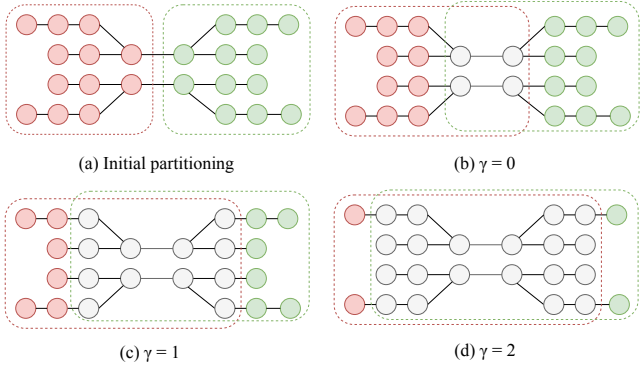

**(a)** Initial partitioning                 **(b)** γ = 0

**(c)** γ = 1                 **(d)** γ = 2

**Figure 4:** *Overlap set computation. The constraint graph is first split using semantic partitioning (a). Nodes at the interface are then identified and placed in the overlap set (b). The set is optionally grown using a user-defined parameter γ in (c) and (d).*

tices and edges may remain fixed for the entirety of the simulation (joints), whereas others might come and go from frame to frame (contacts). Traditional partitioning algorithms such as METIS [KK98] cannot exploit these characteristics so we propose new algorithms for such problems.

We first tackle the issue of creating overlapping sets for an articulated body by processing it offline. Partitioning for freely moving bodies is described in Section 5.4. Consider the bridge in Figure 5: unlike the earlier chain example, it has a fairly complex constraint graph. A key observation is that articulated structures generally have some well-defined *semantic* structure. Ideally, we would want the algorithm to respect the semantics of the articulated mechanism and partition accordingly. For instance, a reasonable partitioning would be to split the bridge along its length into red and green partitions as shown. This is an operation that can be easily performed by the user by dragging a selection box over each portion of the bridge. Thus we rely on the fact that a user may easily intuit and select semantic structures offline and use this partitioning as a starting point to add bodies to the overlap set.

Given this information where the partitioning is approximately balanced, we obtain a constraint graph similar to Figure 4(a). Here, certain edges connect vertices that are in different partitions. We term these *interface edges* and start growing our overlap set from them. The complete procedure is given in Algorithms 2 and 3. The assignment algorithm first iterates through all edges in the constraint graph, searching for an interface edge. When such an edge is found (line 4), both bodies are added to the overlap set in lines 5 and 7 by activating them in the other's workers. Optionally, the set can be grown from here with a user-configurable *growth depth* parameter, referred to as γ. The overlap set is grown from a root body by first gathering all of the root's adjacent neighbors up to depth γ in Algorithm 3 line 2. Then, each neighbor, if not already in the overlap set, is activated in workers that contain the root body in line 5.

The growth depth parameter γ can be seen as a tradeoff between accuracy and speed as demonstrated in Figure 6. A value of γ = 0 will only add the interface bodies to the overlap set, leading to max-imum parallelism but more errors. In contrast, too high a value may result in all bodies being added to the overlap set. A low value of γ introduces errors in the simulation because it results in a poorly conditioned system with large mass ratios. For example, in Figure 6(a) where γ = 0, the whole effective mass of one side of the bridge is transmitted through joint constraints of just four bodies at the interface, leading to large constraint violations as shown. Increasing γ allows more bodies to be shared, distributing the effective mass more evenly, leading to better-conditioned systems at the cost of more shared work. We further refine these errors by calculating non-uniform blending weights, described in Section 5.3.

## 5.2. Blending

We describe the blending algorithm for a single body in this section with pseudocode given in Algorithm 4. The overlap algorithm performs state blending for each body in $\mathcal{O}$ at the end of every main server timestep using state vector **s**, which is a 13-float vector that encodes the rigid transformations and velocities of the body. Consider a body $o$ that is being simulated in $n$ workers. At timestep $t$, worker $i$ calculates its state vector as $\mathbf{s} = (\mathbf{x}_i^t, \mathbf{q}_i^t, \dot{\mathbf{x}}_i^t, \omega_i^t)$. The vectors in $\mathbb{R}^3$ (position, linear velocity, and angular velocity) can be blended in a straightforward manner using a weighted combination

---

**Algorithm 2** Articualted body partitioning.

**Input:** γ is the overlap growth depth
**Input:** $\mathcal{G}$ is the constraint graph of the scene
 1: **function** ASSIGNTOOVERLAPSET(γ, $\mathcal{G}$)
 2:   **for all** joint constraint $c$ in $\mathcal{G}.edges$ **do**
 3:     $A \leftarrow c.bodyA, B \leftarrow c.bodyB$
 4:     **if** $|\mathcal{W}_A| = |\mathcal{W}_B| = 1$ and $\mathcal{W}_A \neq \mathcal{W}_B$ **then**
 5:       $activate(\mathcal{W}_B, A)$ ▷ Add body A to B's workers
 6:       GROWOVERLAP($A$, γ, $\mathcal{W}_A \cup \mathcal{W}_B$, $\mathcal{G}$)
 7:       $activate(\mathcal{W}_A, B)$ ▷ Add body B to A's workers
 8:       GROWOVERLAP($B$, γ, $\mathcal{W}_A \cup \mathcal{W}_B$, $\mathcal{G}$)
 9:     **end if**
10:   **end for**
11: **end function**

---

**Algorithm 3** Grow overlap from a partitioning.

**Input:** $root$ body to grow from
**Input:** $\mathcal{W}_r$ set of workers body $root$ is part of
 1: **function** GROWOVERLAP($root$, γ, $\mathcal{W}_r$, $\mathcal{G}$)
 2:   $\mathcal{N} = \mathcal{G}.getBFSVertices(root, γ)$
 3:   **for all** $b$ in $\mathcal{N}$ **do**
 4:     **if** $b \notin \mathcal{O}$ **then**
 5:       $activate(\mathcal{W}_r \setminus \mathcal{W}_n, b)$
 6:     **end if**
 7:   **end for**
 8: **end function**

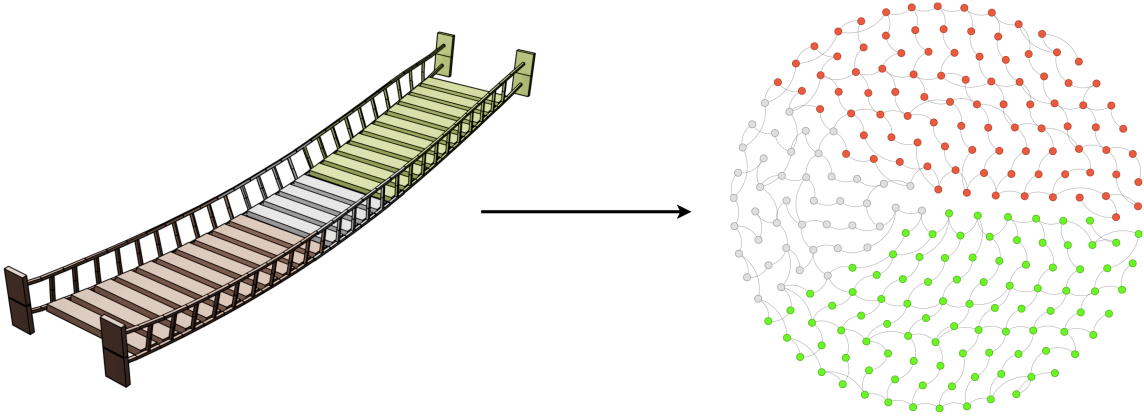

**Figure 5:** *Example of a complex articulated body and its constraint graph after computing overlap set with $\gamma = 2$. Left: bridge composed of 232 rigid bodies with 396 joint constraints connecting them. Red and green colors indicate bodies in different workers and grey indicates bodies in overlap set $\mathcal{O}$. Right: constraint graph $\mathcal{G}$ of the bridge, where each vertex is a rigid body and vertices are connected if they have a joint constraint between them.*

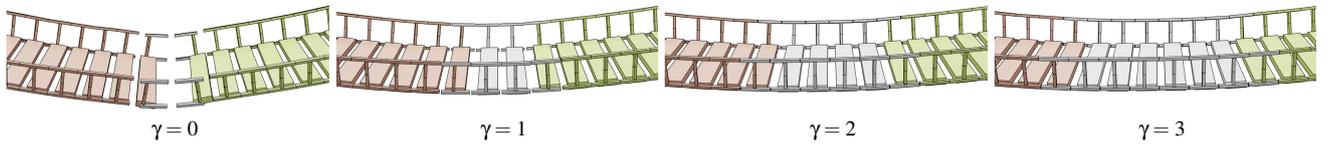

$\gamma = 0$        $\gamma = 1$        $\gamma = 2$        $\gamma = 3$

**Figure 6:** *Cropped view of the same simulation at frame = 70, but with different values of $\gamma$. Increasing $\gamma$ causes the overlap set to grow by including bodies adjacent to the initial four interface bodies and results in a better-conditioned system, meaning gaps between joints get smaller as $\gamma$ is increased.*

as

$$\mathbf{x}_{blended}^t = \sum_{i=1}^{n} w_i \mathbf{x}_i^t, \tag{11}$$

$$\dot{\mathbf{x}}_{blended}^t = \sum_{i=1}^{n} w_i \dot{\mathbf{x}}_i^t, \tag{12}$$

$$\boldsymbol{\omega}_{blended}^t = \sum_{i=1}^{n} w_i \boldsymbol{\omega}_i^t, \tag{13}$$

where the scalar $w_i$ is the convex weight for worker $i$, calculated using the algorithm described in Section 5.3.

For blending rotations, we considered spherical linear interpolation (slerp) [Sho85], but ultimately decided against it as it is expensive to compute and not commutative. Instead, we use a simple naive procedure. We exploit the fact that overlap states are reset at the start of every main timestep, so the rotations calculated by different workers for the same body are always close to each other. This allows us to naively treat the quaternion space as $\mathbb{R}^4$ and simply take a weighted average of each element of the quaternion as

$$\mathbf{q}_{blended}^t = \sum_{i=1}^{n} w_i \mathbf{q}_i^t. \tag{14}$$

However, Equation 14 has two issues [MCCO07] that result in rotation artifacts. The first issue is that quaternions $\mathbf{q}$ and $-\mathbf{q}$ represent the same rotation. Summing such quaternions results in a negatively blended rotation, which is undesirable. We handle

---

**Algorithm 4** Blend algorithm.

**Input:** $o$ is a body in overlap set $\mathcal{O}$
**Input:** $\mathbf{w}$ is the weight vector for body $o$
**Output:** blended state vector $\mathbf{s}$ for body $o$
1: **function** BLEND($o$, $\mathbf{w}$)
2:     $\mathbf{x}_b, \dot{\mathbf{x}}_b, \boldsymbol{\omega}_b \leftarrow (0,0,0); \mathbf{q}_b \leftarrow (0,0,0,0)$
3:     **for** $w_i$ in $\mathbf{w}$ **do**       ▷ For each worker's weight
4:        $\mathbf{x}_b = \mathbf{x}_b + w_i \mathbf{x}_i$
5:        $\dot{\mathbf{x}}_b = \dot{\mathbf{x}}_b + w_i \dot{\mathbf{x}}_i$
6:        $\boldsymbol{\omega}_b = \boldsymbol{\omega}_b + w_i \boldsymbol{\omega}_i$
7:        **if** $i > 0$ and $\mathbf{q}_i . \mathbf{q}_0 < 0$ **then**
8:           $\mathbf{q}_i = -(\mathbf{q}_i^w \mathbf{q}_i^x \mathbf{q}_i^y \mathbf{q}_i^z)$
9:        **end if**
10:       $\mathbf{q}_b = \mathbf{q}_b + w_i \mathbf{q}_i$
11:     **end for**
12:     $\mathbf{q}_b = \|\mathbf{q}_b\|^{-1} \mathbf{q}_b$
13:     **return** $(\mathbf{x}_b, \dot{\mathbf{x}}_b, \boldsymbol{\omega}_b, \mathbf{q}_b)$
14: **end function**

---

this condition in lines 7-10 of the algorithm: when summing each worker's orientation $\mathbf{q}_i^t$ it is compared with the previously accumulated quaternion. If $\mathbf{q}_{i-1}^t = -\mathbf{q}_i^t$ (i.e., their dot product is negative), we flip the sign for each term in $\mathbf{q}_i^t$ which represents the same rotation as $-\mathbf{q}_i^t$ in our case. The second issue is that the resulting blended quaternion is not normalized, which is fixed in line 12.

---

**Algorithm 5** Weight computation.

---
**Input:** $o$ is a body in overlap set $\mathcal{O}$
**Input:** $\mathcal{G}$ is the constraint graph of the scene
**Output:** weights for body $o$
1: **function** COMPUTEWEIGHTS($o$, $\mathcal{G}$)
2:     $d[w] \leftarrow \infty$ for all $w \in \mathcal{W}_o$
3:     $stop[w] \leftarrow False$ for all $w \in \mathcal{W}_o$
4:     **for** $depth$ in $1, 2, ..., \beta$ **do**
5:         $\mathcal{N} \leftarrow \mathcal{G}$.getBFSVertices($o$, $depth$)
6:         **for all** $n$ in $\mathcal{N}$ **do**
7:             **if** $|\mathcal{W}_n| = 1$ and $\mathcal{W}_n \subset \mathcal{W}_o$ **then**
8:                 $d[\mathcal{W}_n] \leftarrow \mathcal{G}$.geodesic($o$, $n$)$^{-1}$
9:                 $stop[\mathcal{W}_n] \leftarrow True$
10:             **end if**
11:         **end for**
12:         **if** all($stop$) **then** break
13:     **end for**
14:     **return** $(\sum d)^{-1} d[w]$ for all $w \in \mathcal{W}_o$
15: **end function**

---

## 5.3. Weight Computation

Intuitively, weights dictate how much a worker's computed state influences the blended state. We define convex weights for an overlap body as

$$\mathbf{w} = (w_1, w_2, ...., w_n), \quad \sum_{i=1}^{n} w_i = 1.0 \text{ and } w_i \geq 0, \qquad (15)$$

where $n$ is the number of workers body $o$ belongs to, $w_i$ is the weight for worker $i$ and the last two conditions enforce convexity. This allows each worker to influence a body's blended state on a spectrum, allowing the use of non-uniform weights when blending state. An interesting observation to note here is that each body in $\mathcal{A}$ has a weight for *every* worker, but some workers implicitly may have a weight of 0. For instance, in Figure 5, red bodies have a weight of 1 for the red worker and 0 for the green worker as it has no influence over a red body's state. In Algorithm 5, we calculate weights only for the workers that have non-zero values.

Now the question is how should these weights be calculated? For simple articulated structures such as the chain in Figure 3, it makes sense for both workers to have an equal say in the body's state (i.e., $\mathbf{w} = (0.5, 0.5)$). However, this naive approach does not work for complex structures. Consider the bridge in Figure 5. If we use equal weights for the bodies in the middle, obvious visual artifacts are seen as the bridge structure oscillates under gravity (Figure 8 bottom right). Motivated by this issue, we now describe a method to compute non-uniform weights, allowing a significant improvement over naive equal weights.

Our main intuition is that some overlap bodies are closer to one worker than the other in constraint graph space, as opposed to standard Euclidean distance proximity. It follows that a worker's influence over a body should be proportional to its graph proximity to said body. For instance, for the bridge, bodies closer to the red worker should be influenced more by it and less so by the green worker. This is a superior method to simply using Euclidean distances, as seen in Figure 7 where we have a large rigid body con-necting two spatially distant bodies. Using constraint graph-based distances allows the long plank to couple spatially distant bricks.

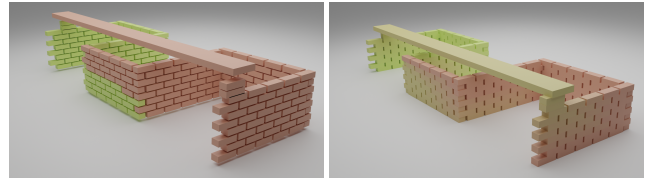

**Figure 7:** *Example of distant coupling. Left: None of the bodies are touching each other. Right: After falling to the ground, geodesic distances allow spatially distant bodies to be coupled by the plank.*

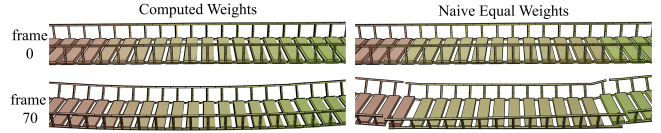

**Figure 8:** *Comparison of computed and naive weights for two workers, red and green. Note the subtle gradient in color in the first column. Computed weights also fix errors produced by naive weights, shown in the second row.*

Thus to quantify a body's distance to a worker in the constraint graph, we use the *graph geodesic distance* $\mathbf{G}(a, b)$ which is defined as the shortest path between vertices $a$ and $b$ in terms of the number of edges. We can view this distance as the number of constraints separating two rigid bodies that are either touching or jointed. The distance of an overlap body $o$ to a worker $i$ is defined as

$$\mathbf{D}(o, i) = \min_{b \, \in \, \mathcal{P}_i \backslash \mathcal{O}} \mathbf{G}(o, b). \qquad (16)$$

In words, $\mathbf{D}(o, i)$ is the geodesic distance of overlap vertex $o$ to the *closest* vertex that is completely in worker $i$. However, simply using the distance metric in this form will result in bodies closer to a worker getting less weight, whereas we want closer bodies to have more weight. Consequently, we use the inverse value $\mathbf{D}(o, i)^{-1}$ to calculate weights.

Pseudocode for this method is given in Algorithm 5. The algorithm performs its search for the closest body completely in one worker by examining neighbors with increasing depths in a breadth-first manner (lines 4-5). The maximum depth for this search is controlled by parameter $\beta$, which is tuned per scene based on how many bodies can be in contact with each other at a given time. This way the search can terminate as soon as a body that meets the criteria is encountered, since that will be the closest body. Values for $\beta$ from 1 to 4 were found suitable for our test scenes. Line 7 ensures the body under consideration belongs to a single worker that is also in this body's overlap set. If true, the inverse distance is recorded in line 8 and a flag indicating search termination for this worker is set in line 9. The search stops when the inverse distance $\mathbf{D}(o, i)^{-1}$ is found for every worker this body be-longs to (line 12). Finally, convex weights proportional to inverse distances are calculated in line 14.

In Figure 8, we visualize weights calculated by this procedure in the left column and naive equal weights in the right column. Unlike earlier examples where overlap bodies were indicated by a grey color, here we manipulate colors to visualize weights: an overlap body's color is derived by blending colors of its base constituent workers per calculated weights. Thus we see all the overlap bodies having the same color in the right column; whereas in the left column a smooth color gradient is achieved, demonstrating the influence of the red worker smoothly fading into the green worker's, based on geodesic proximity. The same is also seen in Figure 7 (right) in the middle part of the wall and areas around where the plank touches the wall.

## 5.4. Dynamic Load Balancing

Contact-rich simulations with a large number of freely moving bodies are unstructured in nature, and can have drastic changes in their constraint graphs from frame to frame due to the dynamic nature of collisions. No obvious semantic partitions exist in these cases, yet there will inevitably be instances where we need to change such assignments. Our key insight here is that if a system starts with an approximately balanced load distribution, we can use inter-worker collision events as opportunities to nudge the distribution towards a more balanced state. We thus start free body simulations by naively dividing all bodies amongst workers and then letting the dynamic load balancing algorithm handle worker assignment during simulation. Naive division here can mean simply dividing the list of rigid bodies equally (see Bowl scene in supplemental video), or selecting areas in a straightforward manner (see Chain Net in supplemental video).

New contacts are obtained by performing a collision detection step on the main server (lines 10-11 in Algorithm 1), as it is aware of the positions of all bodies involved in the simulation. Note that this does not involve solving for constraints or time integration - all we need to know is which bodies are in contact. With this information, it is possible to delineate a finite number of collision scenarios that need to be resolved in the simulation, detailed below. Within the constraint graph, our procedure only evaluates a local neighborhood of the nodes involved in an inter-worker collision event, and the size of this neighborhood can be controlled by the growth depth parameter $\gamma$. Pseudocode for this procedure is in Algorithm 6. We refer the reader to Table 1 for abbreviations and symbols used in the following.

*Inter-worker collision.* A collision involving two bodies that share no common workers is identified as an inter-worker collision in line 4. Figure 9 (i) and (ii) show scenarios handled by this condition: (i) is a collision in isolation whereas in (ii) one of the bodies is already touching another body. Consider scenario (i): no intervention by the main server will cause both bodies to simply pass through each other as each worker is not aware of the other. The goal here is to make one worker aware of the other body, which is done by activating that body in the other worker's (i.e., activate A in $\mathcal{W}_A$ or B in $\mathcal{W}_A$). Note that activating B in the red worker does not imply B's deactivation from the green worker. Rather, B will

---

**Algorithm 6** Load balance algorithm.

**Input:** *contacts* is a list of new contacts
**Input:** *depth* by which to grow overlap set
**Input:** $\mathcal{G}$ is the constraint graph of the scene

1: **function** LOADBALANCE(*contacts*, *depth*, $\mathcal{G}$)
2:     **for all** $c$ in *contacts* **do**
3:         $A \leftarrow c.bodyA, B \leftarrow c.bodyB$
4:         **if** $\mathcal{W}_A \cap \mathcal{W}_B = \varnothing$ **then**     ▷ Figure 9(i),(ii)
5:             **if** $\langle \mathcal{W}_A \rangle \leq \langle \mathcal{W}_B \rangle$ **then**
6:                 *activate*($\mathcal{W}_A$,B)
7:                 *root* $\leftarrow$ B
8:             **else**
9:                 *activate*($\mathcal{W}_B$,A)
10:                *root* $\leftarrow$ A
11:             **end if**
12:             $\mathcal{N} = \mathcal{G}$.getBFSVertices(*root*, *depth*)
13:             **for all** $n$ in $\mathcal{N}$ **do**
14:                *activate*(($\mathcal{W}_A \cup \mathcal{W}_B$) − $\mathcal{W}_{n}$, n)
15:             **end for**
16:         **else if** $|\mathcal{W}_A| > 1$ & $|\mathcal{W}_B| = 1$ & !BRIDGE(A, $\mathcal{G}$) **then**
17:             *deactivate*($\mathcal{W}_A \setminus \mathcal{W}_B$,A)    ▷ Figure 9(iii),(iv)
18:         **else if** $|\mathcal{W}_B| > 1$ & $|\mathcal{W}_A| = 1$ & !BRIDGE(B, $\mathcal{G}$) **then**
19:             *deactivate*($\mathcal{W}_B \setminus \mathcal{W}_A$,B)    ▷ Figure 9(iii), (iv)
20:         **else if** $\mathcal{W}_A = \mathcal{W}_B$ **then**         ▷ Figure 9(v)
21:             $s \leftarrow$ worker with smallest load in $\mathcal{W}_A$
22:             *deactivate*($\mathcal{W}_A \setminus \{s\}$,A)
23:             *deactivate*($\mathcal{W}_B \setminus \{s\}$,B)
24:         **end if**
25:     **end for**
26: **end function**

---

exist in both workers in the next timestep. The question of which body to add where is an opportunity to change load distribution towards a more balanced state. We make this decision by looking at a simple greedy heuristic: the number of bodies in each set at that timestamp (i.e., $\langle \mathcal{W}_A \rangle$ and $\langle \mathcal{W}_B \rangle$), choosing to add to the set with a smaller load and breaking ties arbitrarily. Lines 6-11 implement this logic. We choose to speculatively place bodies in the overlap set here because in practice, it is likely these bodies are near other bodies in the same worker, and they will potentially collide in the upcoming timesteps. In other words, it is very likely scenario (i) will change to (ii) in the next timestep, so preemptively placing one body in the overlap set is justified. If this is not the case, this scenario changes to (iii) in the next timestep, where the overlap body is removed from $\mathcal{O}$.

Now consider scenario (ii): this is essentially the same as (i), but one body (B) is already in contact with another body in the same worker (A). Placing B in the overlap set will allow contact forces from another worker to be transmitted to body A, which is completely in a different worker. We can further enhance this procedure by allowing this overlap set to grow, similar to the procedure in Section 5.1 by using growth depth parameter $\gamma$. This growth is implemented in lines 12-15 and allows us to essentially spread the error across more bodies, reducing visual artifacts.

*Overlap collisions.* Figure 9 (iii), (iv), and (v) show collisions in-

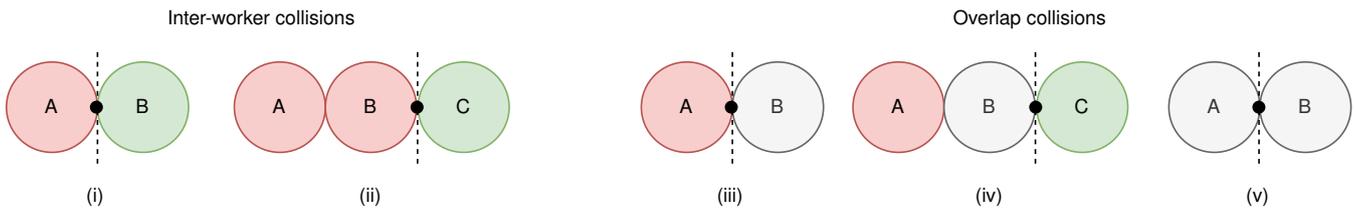

**Figure 9:** *Different scenarios that may arise and cause load rebalancing to trigger on the constraint graph. Here, grey bodies are in the overlap set and coloured bodies belong to a single worker.*

---

**Algorithm 7** Bridge body check.

**Input:** *b* is the body to test
**Input:** $\mathcal{G}$ is the constraint graph
**Output:** *True* if *b* is a bridge body, else *False*

1: **function** BRIDGE($b$, $\mathcal{G}$)
2:      $\mathcal{N} \leftarrow \mathcal{G}$.getAdjNodes($B$), $\mathcal{W} \leftarrow \varnothing$
3:      **for all** $n$ in $\mathcal{N}$ **do**
4:          $\mathcal{W} \leftarrow \mathcal{W} \cup \mathcal{W}_n$
5:      **end for**
6:      **return** $|\mathcal{W}| > 1$
7: **end function**

---

volving bodies in the overlap set, handled in lines 16-24. Consider (iii) and (iv): they both show a body in the overlap set colliding with another body that is in a single worker. In scenario (iii) since B is not touching any other body, it can be safely removed from all workers except the red worker, thus removing redundant computation. However, in scenario (iv) we cannot remove the overlap body B as it is acting as a bridge body between A and C by allowing transmission of contact forces from one worker to the other. This logic is implemented in lines 16-19. Algorithm 7 describes pseudocode to check if a body is acting as a bridge body. Workers of adjacent neighbors of the body are examined: if there is more than one worker, it is a bridge body.

Finally, scenario (v) depicts the case where two overlap bodies collide: this is both an opportunity to remove redundant computation as well as balance load distribution. Using another greedy heuristic, we choose to remove both bodies from all workers except the one with the smallest load in lines 20-24.

## 6. Results

We used the open-source Bullet [CB21] physics simulator in our implementation, which uses a projected Gauss-Seidel iterative solver. The C++ code was modified to allow the activation/deactivation of bodies and optimized API calls to reset states in bulk. Bullet's Python wrapper, PyBullet, was used to implement the main server as it allows rapid prototyping. Additionally, the open-source distributed computing framework Ray [MNW*18] was used to distribute workloads across an Intel i7-9700 and AMD Ryzen 7 5800X both with 16GB RAM, connected over a local network. For frame time measurements, we use the wall-clock time taken for the simulation loop on the main server (lines 6-20 in Algorithm 1). For the baseline, we use the case of a single worker. This means the overlap set $\mathcal{O}$ is always empty, and so the overhead of all related operations is not present, thus essentially running the whole simulation in a single physics engine instance.

**Table 2:** *Test scene information and speedup obtained using a timestep of 2 ms. Speedup is reported for the best-performing number of workers, shown in brackets. The asterisk (\*) shows scenes run in multiple processors on a single CPU.*

| Scene | #Bodies | #Joints | #Contacts (avg) | Speedup |
|---|---|---|---|---|
| **Bowl** | 5040 | 0 | 7022 | 4.10× (7) |
| **Building** | 2408 | 0 | 20471 | 3.27× (8) |
| **Chain Net\*** | 3724 | 0 | 76305 | 5.37× (6) |
| **Four-way Bridge** | 1691 | 3712 | 8924 | 2.86× (3) |
| **Two-way Bridge** | 1798 | 3180 | 3890 | 5.03× (8) |

We use various scenes involving constraints and bodies in the order of thousands to explore the algorithm's scaling, load balancing, and error characteristics. Table 2 shows a summary of scene characteristics and the speedup achieved by the best-performing number of partitions for each scene. We report the average amount of total contacts across the whole simulation, the number of joints, and the number of bodies involved. While Figure 1 shows examples of our different test scenes, we refer the reader to the supplemental video for animations that best illustrate the capabilities and limitations of our method. In all scenes, colors for overlapping bodies are blended according to calculated weights.

*Bowl scene.* We drop 5040 balls in a square bowl with $\gamma = 0$ to explore the load balancing behavior in a scenario where bodies are clumped together. We observe the formation of isolated groups of bodies belonging to single workers, and these are separated by bodies in the overlap set. Due to the greedy nature of heuristics used, we see bodies changing partitions even after settling. In Figure 1, note the green clusters in the center and corners: our method is able to form independent and spatially distant clusters of bodies starting from an initial approximate partitioning.

*Building scene.* We construct a building structure with $\gamma = 0$ and 2408 rigid bodies used for floor, walls, and support columns. We then launch two canon balls to bring it down and observe partitioning. This is an interesting setup to study because unlike the bowl, bodies here are not confined. As soon as the simulation starts, we observe the floors rapidly changing workers because they are simultaneously in contact from the top, bottom, and sides. The bodies that end up off in the distance near the end of the simulation demonstrate the spatially independent nature of our method.

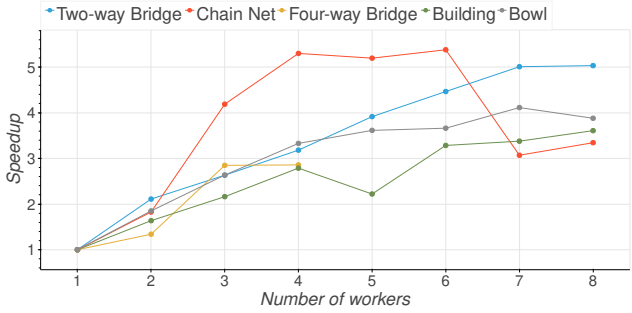

**Figure 10:** *Speedup generally increases as the number of workers is increased for fixed problem sizes.*

*Chain Net scene.* Here we form a net made from 3724 interlocking chain links, each with 1151 vertices. As opposed to primitive shapes used in other scenes, the links produce a very high amount of contacts. This is a challenging scene because the cost of inaccurate solutions is high: if a link tunnels through another, it is immediately noticeable as a hole in the net. Through experimentation we found a value of $\gamma = 3$ to be suitable for this scene. This causes the body assignment to be constantly in flux, avoiding tunneling by spreading the error across multiple links instead of just a few.

*Four-way bridge.* To demonstrate our algorithm's capabilities with overlap bodies that share more than two workers, we designed a structure that comprises two orthogonal bridges connected in the middle on which two balls are dropped to produce interesting dynamics. The bodies in the middle are simulated in either 1, 2, 3, or 4 workers. We found a value of $\gamma = 6$ to be suitable here.

*Two-way bridge.* Similar to Figure 5, we construct a longer bridge with 3180 joint constraints to demonstrate scaling characteristics on articulated bodies. The length of the bridge creates large mass ratios, so we found a value of $\gamma = 6$ as a good compromise between parallelization and constraint errors.

**Scaling analysis.** To evaluate our method's scalability, we ran the test scenes with different numbers of workers. This is an instance of strong scaling analysis, where the problem size is fixed but the number of processors is increased. Figure 11 shows the raw frame times obtained for each scene. For clarity, all plots (except bridge scenes) show averages with a rolling window of size 5. We see frame times for contact-rich scenes (Bowl, Building, and Chain Net) change with the number of contacts present, confirming the assertion that the constraint solve is the dominant step in the simulation loop. To elaborate, frame times for the Bowl and Chain Net scenes increase because more balls/chain links come in contact with each other as the simulation progresses. Conversely, the total number of constraints remain relatively fixed for the two bridge scenes, so frame times remain consistent. We also observe diminishing returns from parallelization as the most gains are made from 2 to 3 workers. It is to be expected that for a fixed problem size, the cost of communication will eventually start to dominate over the benefits of parallelizing the constraint solve. This is clearly seen in the Four-way Bridge plot, where increasing from 3 to 4 workers provides negligible benefit.

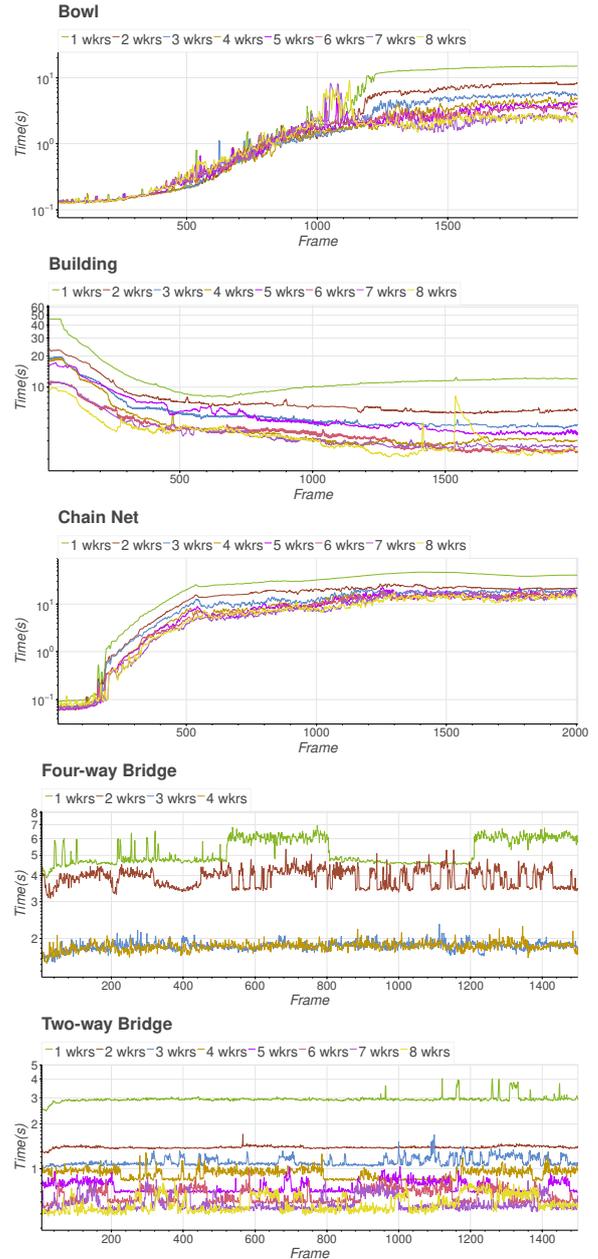

**Figure 11:** *Simulation times for scenes with different numbers of workers. Times decrease as the number of workers is increased.*

However, it is difficult to understand the overall speedup obtained using Figure 11. Instead, Figure 10 clearly shows our algorithm improving simulation performance over the whole duration in the form of a speedup factor. We observe a general trend of improvement in speedup up to a maximum number of workers. The

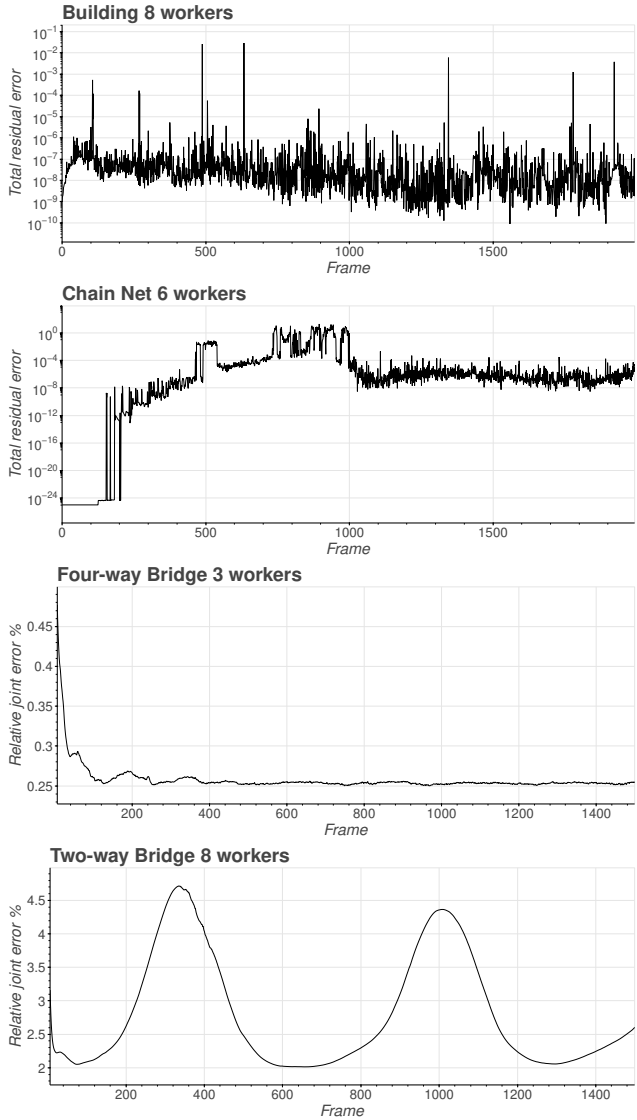

**Figure 12:** *Error plots for best performing number of workers. Errors reach a steady state for the first two scenes despite worker assignment being in flux. The same is seen for the Four-way bridge. The error oscillates for the Two-way Bridge as the structure itself oscillates under gravity.*

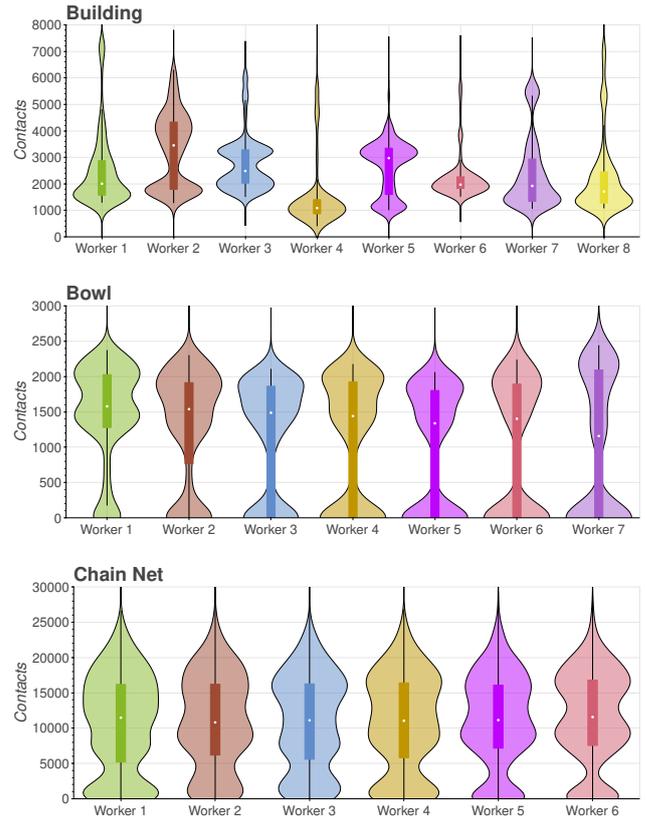

**Figure 13:** *Violin plots of contact distributions. The load balancing algorithm performs well for the Bowl and Chain Net scenes and less so for the Building scene as it contains a disproportionate distribution of contacts per worker.*

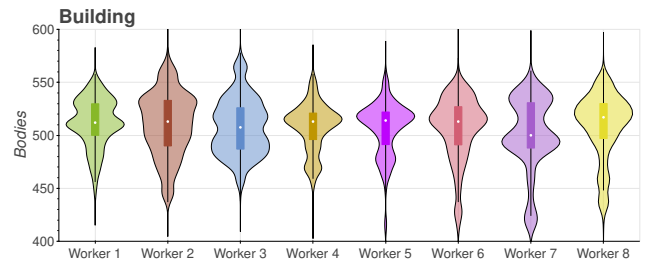

**Figure 14:** *Body distributions for Building scene. The algorithm successfully maintained a balanced distribution of bodies per the heuristic used.*

Two-way bridge is a straightforward example of parallelizing the work equally due to its linear nature, so we see a linear increase in speedup. The Chain Net scene, which contained the most number of constraints, shows the steepest improvement: our algorithm is able to simulate the majority of constraints in parallel while using a minority of overlapping constraints to couple different subsystems. We believe the drop in performance after 6 workers is due to the increase in communication costs.

**Load Balancing analysis.** Since the dominant step is the constraint solve, we examine the distribution of contact constraints within the three contact-rich scenes over the whole duration of the simulation. Figure 13 shows distributions for the best-performing numbers of workers in the form of a violin plot for each worker. The white dot shows the median number of contacts over the whole simulation duration, the bars signify quartiles, and the envelopes show distributions. The Chain Net scene shows well-balanced distributions, in part because the constraint graph topology is fixed.

However, we note that the algorithm maintains a good balance despite (i) the initial partitioning being unequal and (ii) body assignments being constantly in flux (see supplemental video). Likewise, the Bowl scene shows an even distribution of medians, but with larger interquartile ranges for worker 3 and onwards. This is due to the bodies assigned to those workers having no contacts at the start of the simulation. The algorithm achieves this distribution by forming clusters of bodies in spatially distant regions (see different colored clusters in Figure 1), which is something a spatial distribution algorithm would struggle with.

The Building scene highlights an assumption our load balancing heuristic makes, which is that it assumes each body contributes an equal number of contacts. In the simulation, some bodies such as the flat floor pieces have a disproportionately high amount of contacts from all sides. We see the heuristic does its specific job well in Figure 14, which shows an even distribution of bodies across workers. However, the same pattern is not seen in Figure 13, where the workers have an uneven amount of contacts assigned to them. A different heuristic function, such as one that monitors contacts, may improve this result.

**Error analysis.** We direct the reader to view the supplemental video for qualitative results and provide a quantitative report of errors for scenes from the video in Figure 12. For all our test scenes, we set a low residual threshold to force the solver to a maximum of 1000 iterations to produce high quality (low residual) solutions within each worker. This ensures that we can primarily attribute the measured errors to our method. We report the total residual of the LCP system from all workers for Building and Chain Net scenes. We see the residuals increase sharply and then reach a steady state as more contacts are formed and worker assignment evolves. For the two bridge scenes that involve joint constraints, we report position-level total joint violation magnitude with respect to the scene size. Notably, the length of the Two-way bridge causes high effective mass ratios for the portions in the middle, causing gaps to appear. We found a value of $\gamma = 6$ to be acceptable for this case, but note that it can be improved further if desired.

## 7. Limitations and Future Work

Because our algorithm only provides an approximate solution to the full constrained rigid body problem, there are artifacts that can be visible in the results. We rely on the mass of bodies in the overlapping region to provide the necessary coupling forces and the blending of body states in the overlaps permits a smooth distribution of the error. An interesting avenue for future work would be to provide additional procedures for coupling the solutions of different workers, for example, constraint forces from the previous time step [BW97], or by modifying the masses of bodies at simulation boundaries to capture more of the effective mass [PAK*19].

While our work focuses only on dividing the work of the constraint solve, we note that collision detection is a critical part of the simulation that can involve a large amount of computation, particularly in large scenes. One avenue for future work would be to improve performance by distributing the work of collision detection across workers.

Currently, we use a fixed number of workers to perform the simulation that is decided in advance. It would be interesting to modify the method such that new workers are added once existing workers exceed some maximum threshold of work, or removed in cases where there is too little work to distribute. This would allow for simulations with a dynamic number of bodies, for instance, as created or destroyed by fracture or other mechanisms.

Finally, it would be interesting to use different solvers by dividing scenes according to each solver's strengths. For instance, a direct solver could be assigned bodies with large mass ratios and an iterative solver could be used for other bodies.

## 8. Conclusion

We have presented a method to simulate both free and articulated rigid bodies in parallel using heterogeneous CPUs across a network. Our method scales well, allows the user to configure the tradeoff between parallelism and error, and is able to perform load balancing in evolving scenes without the user having to explicitly define any spatial partitioning boundaries. We have demonstrated these aspects in challenging scenarios where physical plausibility is maintained. Additionally, it is straightforward to adapt our method to existing physics engines as it only requires functionality to set/read rigid body states and to activate/deactivate bodies. We imagine that ideas in this work could be useful for permitting large-scale physics simulations to run in real-time on low-end devices such as mobile phones or laptops.